\documentclass[twocolumn,prl,showpacs,amsmath,amssymb]{revtex4}
\usepackage{graphicx}

\begin{document}
%
%
\title{$^3$He Spin-Dependent Cross Sections and Sum Rules}

\author{K.~Slifer,$^{\ntemple}$\setcounter{footnote}{1}\footnote{Presently at University of Virginia}
M.~Amarian,$^{\nyerevan}$
L.~Auerbach,$^{\ntemple}$
T.~Averett,$^{\njlab,\nwm}$
J.~Berthot,$^{\nclermont}$
P.~Bertin,$^{\nclermont}$
B.~Bertozzi,$^{\nmit}$
T.~Black,$^{\nmit}$
E.~Brash,$^{\nregina}$
D.~Brown,$^{\nmaryland}$
% ...
}
\affiliation{
 \baselineskip 2 pt
 \vskip 0.3 cm
 {\rm (Jefferson Lab E94010 Collaboration)} \break
 \vskip 0.1 cm
 \centerline{{$^{\ncaltech}$California Institute of Technology, Pasadena, California 91125}}
 \centerline{{$^{\nchungham}$Chungnam National University, Taejon 305-764, Korea}}
 \centerline{{$^{\nhampton}$Hampton University, Hampton, Virginia 23668}}
 \centerline{{$^{\nclermont}$LPC IN2P3/CNRS, Universit\'e Blaise Pascal, F--63170 Aubi\`ere Cedex, France}}
\centerline{{$^{\niftp}$Institut f\"ur Theoretische Physik II,Ruhr Universit\"at Bochum, D-44780 Bochum, Germany}}
\centerline{{$^{\ninfnp}$INFN, Sezione di Perugia, 06100, Perugia, Italy}}
 \centerline{{$^{\ninfnr}$INFN, Sezione Roma I, P.le A. Moro 2, I-00185, Roma, Italy.}} 
 \centerline{{$^{\ninfn}$INFN, Sezione Sanit\`a, 00161 Roma, Italy}}
 \centerline{{$^{\ninfntor}$INFN, Sezione Tor Vergata, via della Ricerca Scientifica 1, I 00133 Roma, Italy.}}
 \centerline{{$^{\njlab}$Thomas Jefferson National Accelerator Facility, Newport News, Virginia 23606}}
 \centerline{{$^{\nkent}$Kent State University, Kent, Ohio 44242}}
 \centerline{{$^{\nkentucky}$University of Kentucky, Lexington, Kentucky 40506}}
 \centerline{{$^{\nker}$Institut f\"ur Kernphysik, Forschungszentrum J\"ulich, 52425 J\"ulich, Germany}}
 \centerline{{$^{\nkharkov}$Kharkov Institute of Physics and Technology, Kharkov 310108, Ukraine}}
 \centerline{{$^{\nkit}$ Department of Physics, Faculty of Engineering,}}
\centerline{{Kyushu Institute of Technology,
1-1 Sensuicho, Tobata, Kitakyushu 804-8550, Japan}}
\centerline{{$^{\njag}$ M. Smoluchowski Institute of Physics, Jagiellonian University, PL-30059 Krak\`ow, Poland}}
 \centerline{{$^{\nmaryland}$University of Maryland, College Park, Maryland 20742}}
 \centerline{{$^{\nmit}$Massachusetts Institute of Technology, Cambridge, Massachusetts 02139}}
 \centerline{{$^{\numass}$University of Massachusetts-Amherst, Amherst, Massachusetts 01003}}
 \centerline{{$^{\nunh}$University of New Hampshire, Durham, New Hamphsire 03824}}
 \centerline{{$^{\nodu}$Old Dominion University,  Norfolk, Virginia 23529}}
 \centerline{{$^{\nperu}$ Dipartimento di Fisica, Perugia University}} 
 \centerline{{$^{\nprinceton}$Princeton University, Princeton, New Jersey 08544}}
 \centerline{{$^{\nregina}$University of Regina, Regina, SK S4S 0A2, Canada}}
 \centerline{{$^{\nroma}$ Dipartimento di Fisica, Universita$'$ di Roma ``Tor Vergata'', Rome, Italy}}
 \centerline{{$^{\nrutgers}$Rutgers, The State University of New Jersey, Piscataway, New Jersey 08855}}
 \centerline{{$^{\nsaclay}$CEA Saclay, IRFU/SPhN, 91191 Gif/Yvette, France}}
 \centerline{{$^{\nsyracuse}$Syracuse University, Syracuse, New York 13244}}
 \centerline{{$^{\ntemple}$Temple University, Philadelphia, Pennsylvania 19122}}
 \centerline{{$^{\ntohoku}$Tohoku University, Sendai 980, Japan}}
 \centerline{{$^{\ntunl}$Triangle Universities Nuclear Laboratory, Duke Univ. Durham, NC 27708.}}
 \centerline{{$^{\nuva}$University of Virginia, Charlottesville, Virginia 22904}}
 \centerline{{$^{\nwm}$The College of William and Mary, Williamsburg, Virginia 23187}}
 \centerline{{$^{\nyerevan}$Yerevan Physics Institute, Yerevan 375036, Armenia}}
}
\newcommand{\ncaltech}{1}
\newcommand{\nchungham}{2}
\newcommand{\nhampton}{3}
\newcommand{\nclermont}{4}
\newcommand{\niftp}{5}
\newcommand{\ninfnp}{6}
\newcommand{\ninfnr}{7}  
\newcommand{\ninfn}{8}
\newcommand{\ninfntor}{9}
\newcommand{\njlab}{10}
\newcommand{\nkent}{11}
\newcommand{\nkentucky}{12}
\newcommand{\nker}{13}
\newcommand{\nkharkov}{14}
\newcommand{\nkit}{15}
\newcommand{\njag}{16}
\newcommand{\nmaryland}{17}
\newcommand{\nmit}{18}
\newcommand{\numass}{19}
\newcommand{\nunh}{20}
\newcommand{\nodu}{21}
\newcommand{\nperu}{22} 
\newcommand{\nprinceton}{23}
\newcommand{\nregina}{24}
\newcommand{\nroma}{25}        
\newcommand{\nrutgers}{26}
\newcommand{\nsaclay}{27}
\newcommand{\nsyracuse}{28}
\newcommand{\ntemple}{29}
\newcommand{\ntohoku}{30}
\newcommand{\ntunl}{31}
\newcommand{\nuva}{32}
\newcommand{\nwm}{33}
\newcommand{\nyerevan}{34}
\collaboration{Jefferson Lab E94010 Collaboration}
\noaffiliation

\date{\today}
\begin{abstract}
We present a measurement of the spin-dependent cross sections for the
$\vec{^\textrm{3}\textrm{He}}(\vec{\textrm{e}},\textrm{e}')\textrm{X}$ 
reaction in the quasielastic and resonance regions at four-momentum transfer
$0.1\le Q^2\le 0.9$ GeV$^2$. 
The spin-structure functions have been extracted and used to evaluate the 
nuclear Burkhardt--Cottingham and extended GDH sum rules for the first time.
The data are also compared to an impulse approximation calculation and an
exact three-body Faddeev calculation in the quasielastic region.
\end{abstract}
%
%
%
%
%



\maketitle
In recent years, a large amount of high quality spin-dependent data has become
available from a new generation of inclusive electron scattering
experiments~\cite{Chen:2005td}.  These data enable a deeper
understanding of the theory of strong interactions, quantum chromodynamics (QCD), 
via tests of fundamental sum rule predictions.
These predictions are typically derived from
an effective theory or perturbative expansion of QCD, with the choice of appropriate
implementation depending on the four momentum transfer $Q^2$ of the
interaction.  At low $Q^2$, an effective approach known as chiral perturbation 
theory~\cite{Bernard:2007zu}
($\chi$PT) has been tested by several recent spin-dependent measurements in the simplest
systems~\cite{Bosted:2006gp,Dharmawardane:2006zd,Amarian:2002ar,Amarian:2004tmp,
Amarian:2004yf,Prok:2008ev}, and larger $Q^2$ data provide strict tests of
future lattice QCD calculations.
It's crucially important to evaluate these predictions over a
wide range of $Q^2$ to determine their limitations and range of applicability.

Many famous sum rules have been tested with nucleon data, but 
the assumptions made in deriving these
relations often apply regardless of whether the target is a nucleon or a
nucleus.
For example, the Gerasimov-Drell-Hearn (GDH) sum rule~\cite{GDHSHORT} 
for a target of  spin $\cal{S}$, mass $M$, and anomalous  magnetic moment $\kappa$
reads:
\begin{eqnarray}
\int_{\nu_{\texttt{th}}}^\infty \frac{\sigma_\texttt{A}(\nu) - \sigma_\texttt{P}(\nu)}{\nu}d\nu
= -4\pi^2 \mathcal{S} \alpha\left(\frac{\kappa}{M}\right)^2
\label{GDHSUMRULE}
\end{eqnarray}
Here $\sigma_\texttt{A}$($\sigma_\texttt{P}$) represents the cross
section for absorption of a real photon ($Q^2=0$) of energy $\nu$ 
which is polarized 
anti-parallel (parallel) to the target spin
and $\alpha$ is the fine-structure constant.
The inelastic  threshold is signified by $\nu_{\texttt{th}}$,
which is pion production (photodisintegration) 
for a nucleonic (nuclear) target.
Due to the $1/\nu$-weighting, states with lower invariant mass 
provide the most significant contribution to the sum rule.
The GDH predictions for the neutron and $^3$He
are $-234$ and $-496~\mu$b, respectively.  
To gauge the relative strength of the nuclear contribution 
to Eq.~\ref{GDHSUMRULE}, we divide the $^3$He integral  
into two excitation energy regions.
Region I extends from two-body breakup 
to the pion production threshold,
and region II extends from threshold to $\infty$.
Polarized $^3$He  at first order appears  as a free polarized neutron due to the
spin pairing of the two protons, so 
the contribution from region II should be approximately $-234$ $\mu$b.
Therefore, the contribution from disintegration, which is
the only reaction available to real photons in region I,
is necessarily quite large in order to satisfy the sum rule
prediction for $^3$He.  Similarly, in the case of virtual
photon scattering, Ref.~\cite{Arenhovel:2004yu} indicates
the growing importance of threshold disintegration at low
$Q^2$ for the lightest nuclear systems.

Ji and Osborne~\cite{Ji:1999mr} suggest a generalization 
of the GDH sum rule based on  the relationship between
the forward virtual Compton amplitudes $S_1$ and $S_2$, and the spin-dependent structure
functions $g_1$ and $g_2$.  
Since Eq.~\ref{GDHSUMRULE} is derived from
the dispersion relation for 
$S_1$ at the real photon point,
a generalized sum rule can also be constructed from the same relation
at nonzero $Q^2$.
This leads to a set of $Q^2$-dependent
dispersion relations~\cite{Drechsel:2002ar} for the spin-structure  
functions.
In particular, the dispersion relation for $S_1$ leads to the following 
 extension of the GDH sum rule to virtual-photon scattering:
\begin{eqnarray}
\label{JISUMRULE}
 \overline{\Gamma}_1(Q^2) \equiv \int_0^{1-\epsilon} g_1(x,Q^2) dx = 
  \frac{Q^2}{8}~ \overline{S}_1(0,Q^2)
\end{eqnarray}
The infinitesimal $\epsilon$ ensures that only inelastic contributions
are included, which is indicated by the overbar,
and
$x=Q^2/2M\nu$ is the Bjorken scaling variable.
However, an alternate extension is often presented~\cite{Drechsel:2000ct}:
\begin{eqnarray}
\nonumber
I(Q^2) &=& \int_{\nu_{th}}^\infty \frac{\sigma_A(\nu,Q^2)-\sigma_P(\nu,Q^2)}{\nu} d\nu\\
       &=& 2\int_{\nu_{th}}^\infty \frac{K}{\nu}\frac{\sigma_{TT'}(\nu,Q^2)}{\nu} d\nu
\label{GDHX}
\end{eqnarray}
where $K$ is the virtual photon flux factor. 
The spin-dependent contributions to the inclusive cross section of a spin-1/2
system are contained in $g_1$ and $g_2$,
or equivalently the
cross sections $\sigma_{TT}'$ and $\sigma_{LT}'$,
which are the transverse-transverse and 
longitudinal-transverse 
cross sections relevant to scattering with the target spin aligned with, or perpendicular to, 
the direction of the momentum transfer $\vec{q}$. 

The dispersion relation for $S_2$
leads~\cite{Drechsel:2002ar} to the following super-convergence relation: 
\begin{eqnarray}
\label{BCSUMRULE}
\Gamma_2(Q^2) \equiv \int_0^1 g_2(x,Q^2) dx = 0
\end{eqnarray}
which is the Burkhardt--Cottingham (BC) sum rule~\cite{Burkhardt:1970ti}.
The derivation of the BC sum rule 
depends on the convergence of the integral, and 
assumes that $g_2$ is a well-behaved function~\cite{Jaffe:1991qh} 
as $x\to 0$. 

This Letter details a test of the sum rules described above via an inclusive
cross-section measurement in the quasielastic (QE) and resonance regions.
The experiment was performed in  Hall A~\cite{Alcorn:2004sb} 
of the Thomas Jefferson National Accelerator Facility (JLab).
Longitudinally polarized electrons at six incident energies
(0.9, 1.7, 2.6, 3.4, 4.3, and 5.1 GeV)
were scattered from a high-density polarized $^3$He target.
Longitudinal and transverse target polarizations were maintained, allowing
a precision determination of both 
 $g_1^{^\texttt{3}\texttt{He}}(x,Q^2)$
   and 
 $g_2^{^\texttt{3}\texttt{He}}(x,Q^2)$ or alternatively
$\sigma_{TT'}(\nu,Q^2)$ and $\sigma_{LT'}(\nu,Q^2)$.
Full experimental details can be found in 
Refs.~\cite{Amarian:2002ar,Amarian:2004tmp,Amarian:2004yf} 

\begin{figure}
\includegraphics[width=0.47\textwidth]{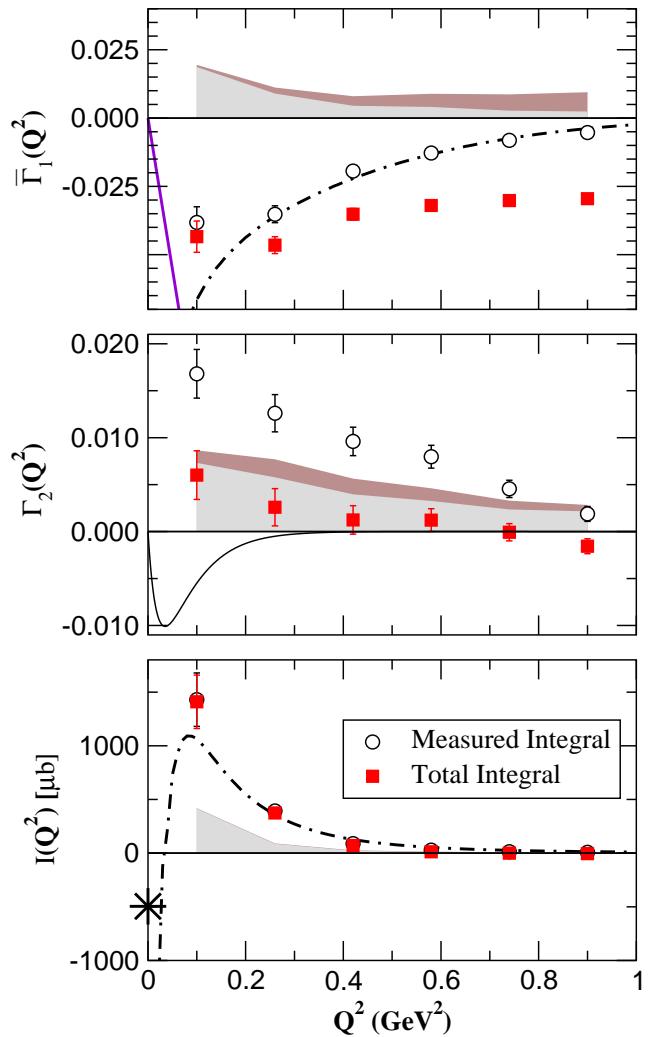}
\caption{\label{3GAM1}  $^3$He spin-structure moments. {\bf Top} : $\overline{\Gamma}_1(Q^2)$
compared to the PWIA model described in text (dot-dash),
and the GDH sum rule slope (solid).
{\bf Middle} : $\Gamma_2(Q^2)$ along with the elastic contribution~\cite{Amroun:1994qj} (solid
) to
the moment.
{\bf Bottom} : $I(Q^2)$ with $K=\nu$, 
compared to the PWIA model.}
\end{figure}

The measured spin-structure functions were interpolated
(or extrapolated for a few data points at large $\nu$)  
to constant $Q^2$~\cite{Amarian:2002ar} from 0.1 to 0.9 GeV$^2$.
Figure~\ref{3GAM1} displays the first moments of 
$g_1^{^3\mathrm{He}}$ and $g_2^{^3\mathrm{He}}$, 
along with the extended GDH sum $I(Q^2)$.  In all panels the circles represent  the
$^3$He data integrated to $W=2$ GeV.
The invariant mass $W$ is defined here in terms of the proton mass: 
$W^2 = {M_p^2 -Q^2 + 2 M_p \nu}$.
Squares include an estimate (discussed below) of any unmeasured contributions.
Statistical uncertainties are shown on the data points, while the systematic uncertainty
of the measured (total) integral is represented by the light (dark) band.
The absolute cross sections contribute 5\% uncertainty, while the
beam and target polarization each contribute 4\%.
The radiative corrections are assigned 20\% 
uncertainty 
to reflect the
variation seen from choosing different initial models for our unfolding procedure.
This uncertainty is doubled for the 0.9 GeV incident energy spectra to reflect 
the lack of lower energy data.  A separate contribution to the radiative
corrections uncertainty arises from the subtraction of the $^3$He elastic radiative
tail, which is significant only for the lowest incident energy. 
The light  band represents the quadratic sum of the above errors including
a contribution from interpolation.
The full systematic band includes an estimate of the uncertainty of the unmeasured 
contribution to the integrals. 
The $\Gamma_2$  full systematic error includes
a 5\% uncertainty from the elastic contribution
(solid black curve) which was evaluated using
previously measured
form factors~\cite{Amroun:1994qj}.

The total integral of $\overline{\Gamma}_1$  includes an 
estimate~\cite{Thomas:2000pf} of the  unmeasured region above $W=2$ GeV,
and the uncertainty arising from this is reflected in the total error band.
Ref.~\cite{Thomas:2000pf} was shown in to be
consistent with existing deep inelastic scattering data~\cite{Airapetian:2002wd} in our previous publication~\cite{Amarian:2004tmp}.
The data show some hint of a turnover at low $Q^2$,
where we have also plotted the slope  predicted by Eq.~\ref{GDHSUMRULE}
for $^3$He.
To obtain the dotted-dashed curve, we have  summed the MAID model~\cite{Drechsel:1998hk} proton and
neutron predictions using an effective polarization
procedure~\cite{CiofidegliAtti:1996cg}.
To this we add an estimate of the contribution below the pion threshold
using the plane-wave impulse approximation (PWIA) model~\cite{CiofidegliAtti:1994cm,Pace:2001cm}.
This model contains contributions for $W\le 2$ GeV, so it should be compared directly
with the open symbols. 
At large momentum transfer, $\Gamma_1(Q^2)$, appears to be nearly independent of $Q^2$,
which would seem to indicate the diminishing importance of higher twist effects,
consistent with other recent findings
(eg. \cite{Meziani:2004ne,Deur:2004ti})  in this kinematic range.

\begin{figure}
\includegraphics[width=8.1cm,height=10.5cm]{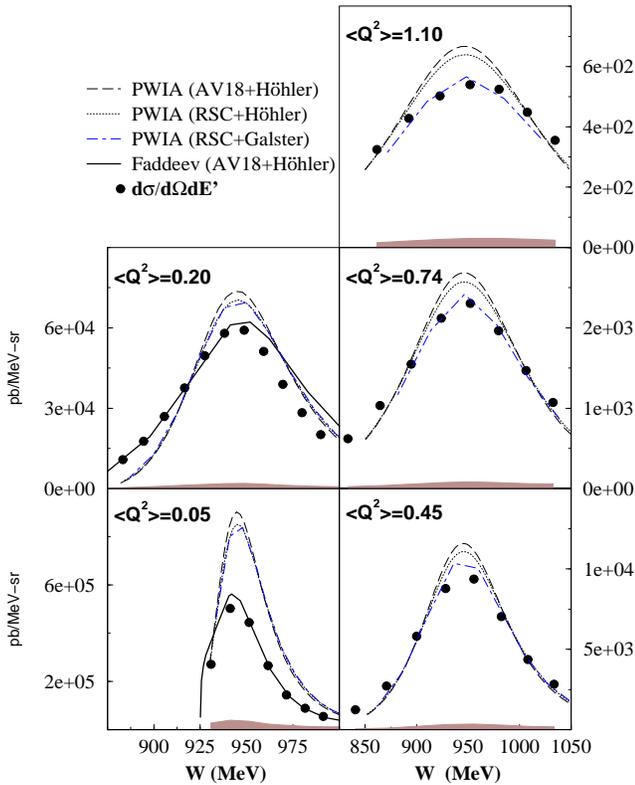}
 \caption{\label{XSALL} $^3$He unpolarized cross sections in the QE region  compared to
  PWIA~\cite{CiofidegliAtti:1994cm,Pace:2001cm} with AV18 (dashed) or RSC (dotted,dot-dashed)
  potential and to the Faddeev calculation~\cite{Golak:2005iy} (solid).
  The error bars (bands) represent statistical (systematic) uncertainties.
  $\langle Q^2 \rangle$ in GeV$^2$.
 }
\end{figure}

\begin{figure}
\includegraphics[width=8.1cm,height=10.5cm]{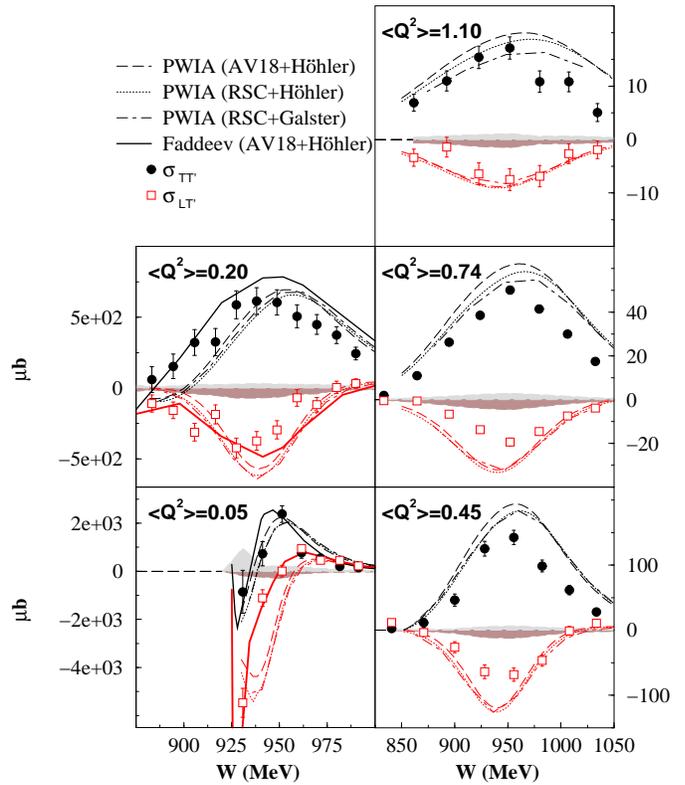}
 \caption{\label{SFALL}  $^3$He polarized cross sections in the QE region.
Curves and notations are the same as in Fig.~\ref{XSALL}.
 }
\end{figure}

Experimental measurements of $g_2$ are scarce and only recently has
the BC sum rule been evaluated for the first time.
The SLAC E155 collaboration~\cite{Anthony:2002hy} measured $\Gamma_2$
at $Q^2=5$ GeV$^2$.  They found the BC sum
rule to be satisfied within a large uncertainty for the deuteron, 
while a violation of almost 3$\sigma$ 
was found for the more precise proton measurement.
In Figure~\ref{3GAM1} (middle)  we plot $\Gamma_2$.  
The unmeasured contribution was estimated using the method described
in~\cite{Anthony:2002hy}, which  assumes the validity of the
Wandzura-Wilczek relation~\cite{Wandzura:1977qf}.
All six data points are consistent with the Burkhardt--Cottingham
prediction.
Results from this same experiment have been used to test the BC 
sum rule for the neutron~\cite{Amarian:2004tmp}, using only data
for which $W>1.073$ GeV, and with nuclear corrections applied.
It was found that the neutron BC sum rule is
satisfied primarily due to the cancellation of the resonance and nucleon elastic
contributions.  It is interesting to find that for $^3$He a balance is struck 
between the 
positive inelastic contribution 
above the pion threshold, and the negative contribution
from the elastic and quasielastic regions, with 
the elastic becoming important  below $Q^2 \approx 0.2$ GeV$^2$.

Figure~\ref{3GAM1} (bottom)  displays the
extended GDH sum as defined in Eq.~\ref{GDHX}.  We follow the
convention $K=\nu$ for the virtual photon flux.
Accounting for the unmeasured contribution~\cite{Thomas:2000pf}
has only a minor effect due to the $1/\nu$-weighting of the integrand.
The phenomenological model (dot-dashed curve)
tracks the data well, but
the negative sum rule prediction at $Q^2 = 0$ (black star) stands in
contrast to the large positive value of our lowest point.
The $^3$He GDH integral is dominated by a
positive QE contribution which largely outweighs the
negative contribution of the resonances.
Assuming the continuity of the integrand 
as $Q^2\to 0$,
as in the nucleonic case~\cite{Arenhovel:1978iy},
our results indicate the necessity of a dramatic turnover in
$I(Q^2)$ at very low $Q^2$. The only possible reaction channel
available to accommodate such a turnover is electro-disintegration at threshold.
Indeed, our 
$\sigma'_{TT}$ 
data~\cite{MYTHESIS} shows an indication of 
a growing negative contribution to the sum in the threshold region as $Q^2$ approaches zero.
A recently completed experiment~\cite{SMANGLE}
may shed light on this behavior.

We focus now on the quasielastic region, where
$^3$He
can  be treated with exact nonrelativistic Faddeev calculations.  This approach
describes existing 
data~\cite{Spaltro:1998,Xu:2000xw,Xiong:2001vb} well at low $Q^2$.
At larger $Q^2$, modern applications~\cite{CiofidegliAtti:1994cm,Pace:2001cm} of 
the PWIA 
have had good success reproducing data.
The measured quasielastic differential cross section ${d\sigma}/{d\Omega dE'}$ 
is displayed in Figure~\ref{XSALL} as a function of W.
In addition, Figure~\ref{SFALL} 
displays the $^3$He polarized cross sections $\sigma_{TT'}$ and $\sigma_{LT'}$.
Radiative corrections have been applied to the data as discussed 
in Refs.~\cite{MYTHESIS,Amarian:2002ar}.
The data are compared to a 
PWIA calculation~\cite{CiofidegliAtti:1994cm,Pace:2001cm} 
and an exact nonrelativistic Faddeev 
calculation~\cite{Golak:2005iy,Xu:2000xw,Xiong:2001vb}. 
The latter includes both final state interactions (FSI) and meson exchange currents (MEC). 
Both groups utilize the H\"ohler~\cite{Howler} 
parameterization for the single 
nucleon current, and the AV18~\cite{Wiringa:1994wb} nucleon-nucleon potential.   
We also display the PWIA curves that
result when the RSC~\cite{CiofidegliAtti:1994cm,Pace:2001cm} potential is used 
instead of AV18, or if the Galster~\cite{Galster:1971kv} form factor parameterization is used instead
of H\"ohler.
In the Faddeev calculation,
the three-nucleon current operator consists of the single
nucleon current and the $\pi$- and $\rho$-like meson exchange
contributions consistent with AV18. 

The Faddeev calculation does not address relativistic effects, and as such
was only performed for the lowest $Q^2$ data. 
The agreement with data is in general quite good, but we find a small discrepancy 
from the data on the high energy side of the 
QE peak at $\langle Q^2 \rangle =0.2$ GeV$^2$.
This  may indicate the increasing importance of 
relativistic effects, along with the growth in relative strength of the $\Delta$ resonance
tail in the QE region as $Q^2$ increases.
We note that $\sigma_{LT'}$ which is not sensitive to the
$\Delta$ resonance generally shows better agreement with the Faddeev calculation
on the high energy side of the quasielastic peak.

At very low $Q^2$, the PWIA calculation fails, 
but improves as expected with increasing momentum transfer,
in part because it takes the relativistic kinematics into account.
The fact that the Faddeev and PWIA calculations differ less as $Q^2$ increases 
seems to indicate that FSI and MEC (neglected in the PWIA) become less
important for these observables as $Q^2$ increases.  
It also appears that the 
PWIA calculation is more sensitive to the choice of the form factor
parameterization, than to the nucleon-nucleon potential utilized.

The Faddeev calculation reproduces the polarized data well at the lowest
$Q^2$, and the PWIA does well at the highest, but there remains an intermediate
zone where both approaches are unsatisfactory.
Refs.~\cite{Xu:2000xw,Xiong:2001vb} 
previously reported that this same PWIA calculation reproduced well
the measured $^3$He quasielastic asymmetry $A_T'$ in this kinematic region. 
As such, we compared this calculation directly to the transverse asymmetry $A_T'$ data 
from our 
experiment and found good agreement, consistent with the previous results
but only in a narrow window centered on the QE peak.

To summarize, we find the Burkhardt--Cottingham sum rule to hold for $^3$He.
The GDH integral and $\Gamma_1$ display intriguing behavior at low $Q^2$,
and will provide valuable constraints on future $\chi$PT and lattice 
QCD calculations.
We have measured the first precision polarized  cross-sections
in the quasielastic and resonance regions of $^3$He. 
A full three-body Faddeev  calculation agrees well with the data, 
but starts to exhibit discrepancies as the energy increases, 
possibly due to growing relativistic effects.
As the momentum transfer increases, the PWIA approach
reproduces the data well, but there exists an intermediate range where
neither calculation succeeds.

\begin{acknowledgments}
This work was supported by the U.S. Department of Energy, 
the National Science Foundation, the French Commissariat 
\`a l'Energie Atomique (CEA),
the Centre National de la Recherche Scientifique (CNRS) and by
the Helmholtz Association through the virtual institute ``Spin and strong QCD''
(VH-VI-231). The numerical calculations were performed on the IBM Regatta p690+
of the NIC in J\"ulich, Germany.
The Southeastern Universities 
Research Association  operates the Thomas Jefferson National Accelerator 
Facility for the DOE under contract DE-AC05-84ER40150, mod. \# 175.
\end{acknowledgments}

%

%
%
%
%
%
%
%
%
%
%
%
%
%
%
%
%
%
%
%
%
%
%
%
%
%
%
%
%
%
%
%
%
%
%
%
%
%
%
%
%
%
%
%
%
%
%
%
%
%
%
%
%
%
%
%
%
%
%
%
%
%
%
%
%
%
%
%
%
%
%
%
%
%
%
%
%
%
%
%
%
%
%
%
%
%
%
%
%
%
%
%
%
%
%
%
%
%
%
%
%
%
%
%
%
%
%
%
%
%
%
%
%
%
%
%
%
%
%
%
%
%
%
%
%
%
%
%
%
%
%
%
%
%
%
%
%
%
%
%
%
%
%
%
%
%
%
%
%
%
%
%
%
%
%
%
%
%
%
%
%
%
%
%
%
%
%
%
%
%
%
%
%
%
%
%
%
%
%
%
%
%
%
%
%
%
%
%
%
%
%
%
%
%
%
%
%
%
%
%
%
%
%
%
%
%
%
%
%
%
%
%
%
%
%
%
%
%
%
%
%
%
%
%
%
%
%
%
%
%
%
%
%
%
%
%
%
%
%
%
%
%
%
%
%
%
%
%
%
%
%
%
%
%
%
%
%
%
%
%
%
%
%
%
%
%
%
%
%
%
%
%
%
%
%
%
%
%
%
%


\end{document}